\documentclass{appolb}
\usepackage{epsfig,cite,amsmath,amssymb}
\newcommand{\ppbar}{p^{\!\!\!\!\!\textsuperscript{\tiny{(--)}}}\!\!}

\begin{document}
\title{Looking for signals beyond the neutrino Standard Model
\thanks{Presented by F. del Aguila at the XXXI International School of 
Theoretical Physics ``Matter To The Deepest: Recent Developments in 
Theory of Fundamental Interactions", Ustro\'n, Poland, September 5-11, 
2007.}%
}
\author{F. del Aguila, J.A. Aguilar Saavedra, J. de Blas
\address{Departamento de F{\'\i}sica Te\'orica y del Cosmos and CAFPE, \\
Universidad de Granada, E-18071 Granada, Spain}
\and
M. Zra{\l}ek
\address{Department of Field Theory and Particle Physics, Institute of Physics, \\
University of Silesia, Uniwersytecka 4, 40-007 Katowice, Poland}
%
}
\maketitle
\begin{abstract}
Any new neutrino physics at the TeV scale must include a suppression mechanism
to keep its contribution to light neutrino masses small enough.
We review some seesaw model examples with weakly broken lepton number,
and comment on the expected effects at 
large colliders and in neutrino oscillations.
\end{abstract}
\PACS{14.60.St,13.35.Hb,14.60.Pq,13.15.+g}


\section{Introduction}

Lepton flavour-changing processes have been only observed in neutrino 
oscillations \cite{Yao:2006px}.
These can be explained by introducing nonzero neutrino masses 
and the corresponding charged current mixing matrix (MNS) 
\cite{MNS}
which relates neutrino mass and current eigenstates. This defines the minimal 
neutrino Standard Model $\nu$SM \cite{Mohapatra:1998rq},
which can be realised with the
addition of a Majorana mass term or introducing three light right-handed
neutrinos with Yukawa couplings to the SM ones.

The small size of the light neutrino masses, $m_{\nu_i} \sim 1$ eV, makes
the observation of neutrino mixing very difficult.
In neutrino oscillations the long baseline distance $L$ enhances the small
ratio $\Delta m^2_{ij} / E_\nu$, where $\Delta m^2_{ij} = m^2_{\nu_i} -
m^2_{\nu_j}$, making the relevant quantity for neutrino oscillations
\begin{equation}
\frac{\Delta m^2_{ij} [{\rm eV}^2] \; L [{\rm km}]}
{E_{\nu} [{\rm GeV}]} 
\end{equation}
of order unity. However,
in high energy
collider experiments the available luminosities cannot sufficiently enhance the
small mass ratios $m^2_{\nu_i} / E^2$, with $E$ the relevant
energy scale in the process, and then lepton flavour violating (LFV) effects
are negligible. Hence, the observation of lepton flavour violation at colliders 
will imply new physics near the TeV scale, which is the scale to be probed at
LHC. Conversely, it is also expected that if there is new physics at this scale,
it violates lepton flavour because the new interactions do not need to be
aligned with the neutrino current eigenstates in general. 
Any extended model with new neutrino physics near the electroweak scale must
include a mechanism for decoupling the generation of light neutrino masses
from the physics at the new scale. In section 2 we discuss how this works in
the three types of seesaw. The symmetry protecting light neutrino masses
appears to be in all three cases lepton number conservation. We also
discuss
the limits on the coefficients of the dimension 6 operators parameterising the
new physics at low energy. Limits on masses and mixings of heavy neutrinos at
large colliders like ILC, CLIC and LHC are
presented in section 3. Section 4 is devoted to new possible effects in
neutrino oscillations.

\section{Low energy physics}

At energies much smaller than the mass of any new resonance, the departures
from the SM can be
parameterised by an effective lagrangian, which is determined by the
light field content and the required symmetries. The precision is given by the
order considered in the momentum expansion. In the case of light neutrinos
the effective Lagrangian depends on their Dirac or Majorana nature.
In the Dirac case we have to introduce at least three new right-handed
neutrinos to pair with the SM left-handed counterparts, and lepton number is
in principle conserved. The smallness of SM neutrino masses stems from the
smallness of the Yukawa couplings, which requires a satisfactory explanation. In
the Majorana case the field content is the same as 
in the SM, light neutrinos are Majorana particles and lepton
number is broken. The smallness of neutrino masses is related to the
large scale of this symmetry breaking.

We will concentrate on the second possibility.
The most general effective lagrangian invariant under 
$\mathrm{SU}(3)_C \times \mathrm{SU}(2)_L \times \mathrm{U}(1)_Y$,
\begin{equation}
{\mathcal L}^{eff} = {\mathcal L}_4 + \frac{1}{\Lambda} {\mathcal L}_5 
+ \frac{1}{\Lambda^2} {\mathcal L}_6 + \dots \,,
\end{equation}
is explicit up to dimension 6 in Ref.~\cite{Buchmuller:1985jz}. 
${\cal L}_4$ stands for the SM lagrangian, ${\cal L}_5$  contains the only
dimension 5 operator  allowed by gauge
symmetry,\footnote{We use the operator basis of Ref.~\cite{Buchmuller:1985jz}. 
$L$ stands for the lepton doublet and $\tilde{\phi} = i \sigma _2 \phi ^*$ 
is the Higgs doublet with hypercharge $Y = -\frac{1}{2}$. Family indices are
not shown,  unless otherwise stated.}
\begin{equation}
{\mathcal O}_5 = \overline{L^c} \tilde{\phi}^* \tilde{\phi}^\dagger L \ , 
\end{equation}
and ${\mathcal L}_6$ includes all dimension 6 operators (81 without taking
into account flavour indices) which preserve lepton and baryon number. 
This lagrangian is valid for energies below $\Lambda$, the cut-off scale. 
After spontaneous symmetry breaking ${\mathcal O}_5$ generates light neutrino
Majorana masses  $m_\nu = -x_5 v^2 / \Lambda$, being $x_5/\Lambda$ the
coefficient of 
this dimension 5 operator and $v=246$ GeV the Higgs vacuum expectation value.
For $m_\nu \sim 1$ eV, as required by experimental data,
$\Lambda \sim 10^{14}$ GeV if
 $x_5 \sim 1$, or $x_5 \sim 10^{-11}$ if $\Lambda \sim 1$ TeV. In the 
first case new physics cannot manifest itself in any high energy experiment
considered up to now.  In the latter, new effects can 
show up in the new generation of accelerator experiments if the coefficients 
of the dimension 6 operators are relatively large. However, in this scenario 
one has to explain why the coefficient of $\mathcal{O}_5$ is so small. 
The simplest models including such a decoupling mechanism distinguish between 
the cut-off scale $\Lambda$ and the effective lepton number violating (LNV)
parameter entering  in the definition of $x_5$ (see for examples 
Refs.~\cite{Abada:2007ux,Chen:2007dc}). In the rest of this section we partly
review the results in  Refs.~\cite{Abada:2007ux,Kersten:2007vk},
following approximately the notation in Ref.~\cite{Abada:2007ux}. 
The different values of the coefficients reflects the different 
normalisation and the different operator basis used.

The minimal SM extension exhibiting this decoupling only requires the
addition of heavy Dirac neutrino singlets $N$. In this case lepton number can
be assigned so that left-handed fields $N_L$ have quantum number $\mathbf{q}$
and $-\mathbf{q}$ the right-handed counterparts
$N^c_R$. Then, the generic mass matrix reduces to
\begin{equation}
\label{matrix}
\begin{array}{c} \\[0.5mm] \nu_L \\[1mm] N \end{array} \!\!\!\!
\begin{array}{cc}
\ \ \nu_L & N \\[0.2cm]
\left( \!\! \begin{array}{c}
0\\Y_N\frac{v}{\sqrt{2}}\end{array}\right. &\left.\begin{array}{c}
Y^T_N\frac{v}{\sqrt{2}}\\M_N\end{array}\!\!\right)
\end{array}
~ \raisebox{-2.05mm}{$\longrightarrow$} ~
\begin{array}{c}\\ \nu_L\\N_L\\N_R^c\end{array} \!\!\!\!
\begin{array}{c c l}
\ \ \nu_L& N_L& \ \ N_R^c\\[0.2cm]
\left(\begin{array}{c}
0\\0\\\frac{y_N \,v}{\sqrt{2}}\end{array}\right.&\begin{array}{c}
0\\0\\m_N\end{array}&\left.\begin{array}{c}
\frac{y_N \, v}{\sqrt{2}}\\m_N\\0\end{array}\right)
\end{array}
\end{equation}
for one family,
where $y_N$ is the Yukawa coupling between the SM neutrino and the
right-handed one. If $y_N \neq 0$, $N_L$ and $\nu _L$ have the same
lepton number, $\mathbf{q} = 1$,
and they mix. When lepton number is broken by a small entry $\mu$ instead of
some of
the zeroes in the above matrix, the light neutrino gets a Majorana mass
proportional to it, even if the nonzero entry is in the $(3,3)$ position
because one-loop radiative corrections also generate a nonzero mass
for $\nu _L$ proportional to $\mu$. (A similar behaviour is found in Little
Higgs models \cite{del Aguila:2005yi}.)

More generally, all three types of seesaw mechanisms generating 
${\cal O}_5$ at the tree level upon integration of heavy fields 
\cite{Arzt:1994gp} can incorporate a similar decoupling. 
In Tables \ref{Fsinglet}--\ref{Ftriplet} we collect the operators 
up to dimension 6 obtained from the integration of 
heavy fermion singlets $N$ (type I seesaw), 
scalar triplets $\Delta$ (type II) and 
fermion triplets $\Sigma$ (type III), 
respectively, and the corresponding coefficients \cite{Abada:2007ux},
where now $\Lambda$ is the mass of the heavy resonance. 
\begin{table}[htb]
\begin{center}
\begin{tabular}{| c | c l | c |} \hline
$\mbox{Dimension}$&$ $&$\quad \quad\mbox{Operator}$&$\mbox{Coefficient}$\\
\hline\hline
~&~&~\\[-0.4cm]
$5$&${\cal O}_{5}$&$= \overline{L^c} \tilde \phi^* \tilde \phi^\dagger L$&
$\frac 12 Y_N^T M_{N}^{-1} Y_N$\\[1mm]
\hline
~&~&~\\[-0.4cm]
$6$&${\cal O}_{\phi L}^{(1)}$&$=\left(\phi^{\dagger}iD_\mu \phi\right)
\left(\overline{L} \gamma^\mu L\right)$&$\frac 14 Y_N^\dagger
(M^\dagger_{N})^{-1} M_N^{-1} Y_N$\\[1mm]
$ $&${\cal O}_{\phi L}^{(3)}$&$=\left(\phi^{\dagger}i\sigma_a D_\mu \phi\right)
\left(\overline{L} \sigma_a \gamma^\mu L\right)$&$-\frac 14 Y_N^\dagger
(M^\dagger_{N})^{-1} M_N^{-1} Y_N$\\[1mm]
\hline
\end{tabular}
\caption{Operators arising from the integration of heavy Majorana fermion singlets 
$N$. $Y_N$ is the coupling matrix in the Yukawa term 
$-\overline{L} \tilde{\phi} Y_N^\dagger N_R$.}
\label{Fsinglet}
\end{center}
\end{table}
\begin{table}[htb]
\begin{center}
\begin{tabular}{| c | c l | c |} \hline
$\mbox{Dimension}$&$ $&$\quad \quad\mbox{Operator}$&$\mbox{Coefficient}$\\
\hline\hline
~&~&~\\[-0.4cm]
$4$&${\cal O}_{4}$&$= \left(\phi^\dagger \phi\right)^2$&
$2 \left|\mu_\Delta\right|^2 / M_{\Delta}^2$\\[1mm]
\hline
~&~&~\\[-0.4cm]
$5$&${\cal O}_{5}$&$= \overline{L^c} \tilde \phi^* \tilde\phi^\dagger L$&
$-2 \, Y_\Delta \mu_\Delta / M_{\Delta}^2$\\[1mm]
\hline
~&~&~\\[-0.4cm]
$6$&${\cal O}_{LL}^{(1)}$&$=\frac 12\left(\overline{L^i} \gamma^\mu L^j\right)
\left(\overline{L^k} \gamma_\mu L^l\right)$&$2 / M^2_{\Delta}
(Y_\Delta)_{jl} (Y_\Delta^\dagger)_{ki}$\\[1mm]
$ $&${\cal O}_{\phi}$&$=\frac 13 \left(\phi^{\dagger}\phi\right)^3$&
$-6\left(\lambda_3+\lambda_5\right)
\left|\mu_\Delta\right|^2 / M^4_{\Delta}$\\[1mm]
$ $&${\cal O}_{\phi}^{(1)}$&$=\left(\phi^{\dagger}\phi\right)
\left(D_\mu \phi\right)^\dagger D^\mu \phi$&
$4 \left|\mu_\Delta\right|^2 / M^4_{\Delta}$\\[1mm]
$ $&${\cal O}_{\phi}^{(3)}$&$=\left(\phi^{\dagger}D_\mu \phi\right)
\left( D^\mu \phi^\dagger \phi\right)$&
$4 \left|\mu_\Delta\right|^2 / M^4_{\Delta} $\\[1mm]
\hline
\end{tabular}
\caption{Operators arising from the integration of heavy scalar triplets 
$\Delta$. $Y_\Delta$ is the coupling matrix in the Yukawa term 
$\overline{\tilde{L}} Y_\Delta \left(\vec{\sigma} \cdot \vec{\Delta}\right) L$, 
with $\overline{\tilde{L}} = -L^T C i \sigma _2$ and $C$ the matrix entering the 
spinor charge conjugation definition;  
and $\mu_\Delta$, $\lambda_3$ and $\lambda_5$ are the coefficients of the 
scalar potential terms 
$\tilde \phi^\dagger\left(\vec{\sigma}\cdot \vec{\Delta}\right)^\dagger \phi$, 
$-\left(\phi^\dagger \phi\right)\left(\vec{\Delta}^\dagger \vec{\Delta}\right)$ 
and 
$-\left(\vec{\Delta}^\dagger T_i \vec{\Delta}\right)\phi^\dagger \sigma_i \phi$, 
respectively.}
\label{Striplet}
\end{center}
\end{table}
\begin{table}[htb]
\begin{center}
\begin{tabular}{| c | c l | c |} \hline
$\mbox{Dimension}$&$ $&$\quad \quad\mbox{Operator}$&$\mbox{Coefficient}$\\
\hline\hline
~&~&~\\[-0.4cm]
$5$&${\cal O}_{5}$&$= \overline{L^c} \tilde{\phi}^* \tilde{\phi}^\dagger L$&
$\frac{1}{2} Y_\Sigma^T M_{\Sigma}^{-1} Y_\Sigma$\\[1mm]
\hline
~&~&~\\[-0.4cm]
$6$&${\cal O}_{\phi L}^{(1)}$&$=\left(\phi^{\dagger}iD_\mu \phi\right)
\left(\overline{L} \gamma^\mu L\right)$&$\frac{3}{4} Y_\Sigma^\dagger
(M^\dagger_{\Sigma})^{-1} M_\Sigma^{-1} Y_\Sigma$\\[1mm]
$ $&${\cal O}_{\phi L}^{(3)}$&$=\left(\phi^{\dagger}i\sigma_a D_\mu \phi\right)
\left(\overline{L} \sigma_a \gamma^\mu L\right)$&$\frac{1}{4}
 Y_\Sigma^\dagger
(M^\dagger_{\Sigma})^{-1} M_\Sigma^{-1} Y_\Sigma$\\[1mm]
$ $&${\cal O}_{l \phi}$&$=\left(\phi^{\dagger}\phi\right)\overline{L} \phi l_R$&
$Y_\Sigma^\dagger (M^\dagger_{\Sigma})^{-1} M_\Sigma^{-1} Y_\Sigma Y_l$\\[1mm]
\hline
\end{tabular}
\caption{Operators arising from the integration of heavy Majorana fermion triplets 
$\Sigma$. $Y_\Sigma$ is the coupling matrix in the Yukawa term  
$-\overline{\vec{\Sigma}_R} Y_\Sigma (\tilde{\phi}^\dagger \vec{\sigma} L)$
and $Y_l$ in $-\overline{L} Y_l \phi l_R$.}
\label{Ftriplet}
\end{center}
\end{table}
In type II seesaw the coefficient of ${\cal O}_5$ is 
explicitly proportional to the LNV product 
$\mu_\Delta Y_\Delta$, while none of the other coefficients contains 
both parameters. This allows for a relatively light
scalar triplet with
$M_\Delta \sim 1$ TeV and possibly observable effects at forthcoming
experiments, while keeping SM neutrino masses very small (in definite models
\cite{Chen:2007dc} there can be also extra loop suppression factors). 
In the other two types of seesaw the decoupling is not so explicit.
In both cases the coefficient of the dimension 5 operator is proportional to
$Y^T M^{-1} Y$, thus it only depends (quadratically) 
on $Y$, while the coefficients of the dimension 6 operators
involve $Y^\dagger (M^\dagger)^{-1} M^{-1} Y$, with $Y$ and $Y^\dagger$.
In this way it is possible that there are cancellations in the former product 
which do not hold in the latter one.
This is indeed what happens for quasi-Dirac neutrinos. 
For our one-family example 
in Eq. (\ref{matrix}), if the LNV parameter $\mu$ is in the (2,2) position,
the SM neutrino acquires a Majorana mass $m_\nu$ \footnote{The
$2 \times 2$ bottom-right submatrix must be diagonalised before applying the
seesaw formula in order to make the cancellation apparent.
The masses of the two Majorana eigenstates are taken to
be positive, $m_{N_1} \simeq m_N + \mu/2$, 
$m_{N_2} \simeq m_N - \mu/2$.} 
\begin{equation}
-  Y_{N}^T M_{N}^{-1} Y_{N}  \frac{v^2}{2}
\simeq - \frac{y_N^2}{2} \left[
\frac{(1-\frac{\mu}{4m_N})^2}{m_N+\frac{\mu}{2}} 
-\frac{(1+\frac{\mu}{4m_N})^2}{m_N-\frac{\mu}{2}}
\right] \frac{v^2}{2}
\simeq \frac{\mu y_N^2}{m_N^2} \frac{v^2}{2},  
\end{equation}
where we only keep the dominant terms in $\mu/m_N$.
(Less natural cancellations are also possible in more involved 
models \cite{Kersten:2007vk,del Aguila:2005mf}.)
While $m_\nu$ is proportional to $\mu$, the coefficients of the dimension 
6 operators are not,
\begin{equation}
Y_{N}^\dagger (M_{N}^\dagger)^{-1} M_{N}^{-1} Y_{N} 
\simeq  \frac{|y_N|^2}{2} \left[ 
\frac{(1-\frac{\mu}{4m_N})^2}{(m_N+\frac{\mu}{2})^2} 
+\frac{(1+\frac{\mu}{4m_N})^2}{(m_N-\frac{\mu}{2})^2}
\right] \simeq \frac{|y_N|^2}{m_N^2} \,. 
\end{equation}
Hence, new fermions can exist near the TeV scale with observable effects
beyond the SM in future experiments, while maintaining the SM neutrinos light
enough.

Present experimental limits on the different combinations
of quadratic products of Yukawa couplings $y^* y$ entering the
dimension 6 operators range from 0.3 to 0.002 for a heavy
neutrino singlet $N$ with a mass of 1 TeV;
from 1 to $10^{-5}$ for a heavy scalar triplet $\Delta$
of the same mass, and from 0.01 to
$3 \times 10^{-5}$ for a heavy fermion triplet $\Sigma$
equally heavy . A detailed analysis can be found in
Ref. \cite{Abada:2007ux}.

\section{Lepton signals at large colliders}

The next generation of large colliders will be able to further constrain the
masses and mixings of the seesaw messengers (see Ref. \cite{del Aguila:2006dx}
for a review in the case of heavy neutrino singlets). Here we restrict
ourselves to $e^+ e^-$ and hadron colliders.

\subsection{$e^+ e^-$ colliders}

\begin{figure}[htb]
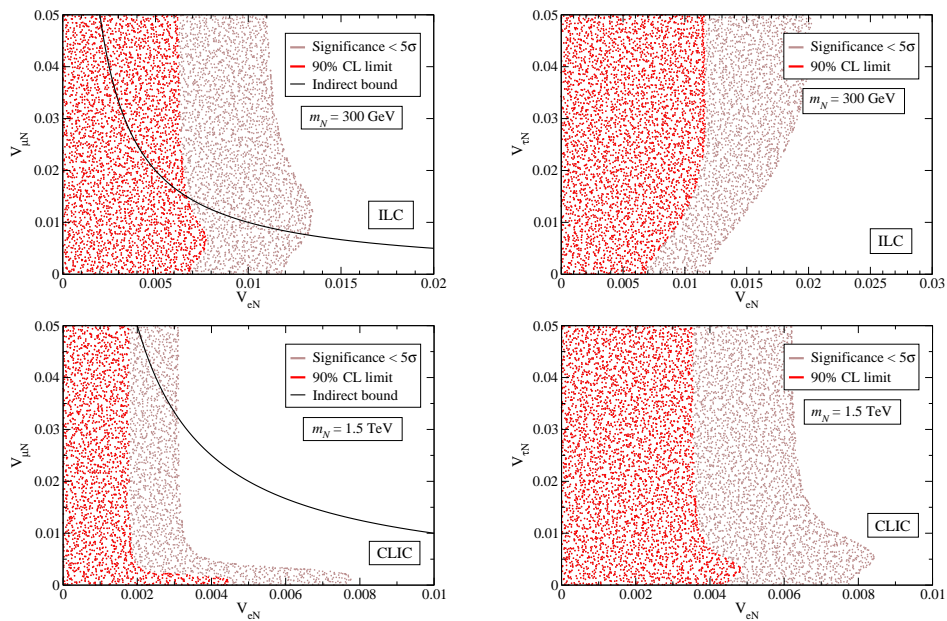

\begin{center}
\begin{tabular}{ccc}
\epsfig{file=Figs/bound-em-ILC.eps,height=4cm,clip=} & &
\epsfig{file=Figs/bound-et-ILC.eps,height=4cm,clip=} \\
\epsfig{file=Figs/bound-em-CLIC.eps,height=4cm,clip=} & &
\epsfig{file=Figs/bound-et-CLIC.eps,height=4cm,clip=}
\end{tabular}
\caption{Combined limits on heavy neutrino mixings at ILC (up) and
CLIC (down), for the cases $V_{\tau N} = 0$ (left) and  
$V_{\mu N} = 0$ (right).
The red areas represent the 90\% CL limits if no signal is observed. 
The white areas extend up to present bounds 
$V_{eN} \leq 0.073$, $V_{\mu N} \leq 0.098$,
$V_{\tau N} \leq 0.13$ \cite{Bergmann:1998rg,Bekman:2002zk}
, and correspond to the region where a combined
statistical significance of $5\sigma$ or larger is achieved. 
The indirect limit from $\mu - e$ LFV processes is also shown.}
\label{fig:ilc}
\end{center}
\end{figure}

The process $e^+e^- \rightarrow N \nu \rightarrow \ell^{\pm} W^{\mp} 
(\rightarrow q \bar q') \nu $
sets the most stringent limits on the mass and the mixing of a heavy 
neutrino singlet (seesaw type I) for a large enough center of mass 
energy so that $N$ is produced \cite{del Aguila:2005pf}
(see also Ref. \cite{Gluza:1996bz}).
Lepton colliders are a rather clean environment, being the irreducible 
background for this process SM four-fermion $\ell \nu q \bar q'$ production
(which includes $W^+ W^-$ plus non-resonant diagrams).
The non-observation of an excess in the $\ell jj$ invariant mass 
distribution will set limits on the heavy neutrino mass $m_N$ and
its mixing with the charged leptons 
$V_{\ell N} = Y^*_{N \ell} v/(\sqrt 2 m_N)$, $\ell = e, \mu, \tau$. 
Limits are rather independent of $m_N$ up to nearly the kinematical 
limit, and independent of the Dirac or Majorana character of the heavy 
neutrino. In Fig.~\ref{fig:ilc} we show the combined limits on the 
mixing of a new heavy neutrino singlet
(i) at ILC, with a centre of mass energy $\sqrt s = 500$ GeV
and an integrated luminosity $L = 345$ fb$^{-1}$,
taking $m_N = 300$ GeV;
(ii) at CLIC, with $\sqrt s = 3$ TeV, $L = 1000$ fb$^{-1}$, and
taking $m_N = 1.5$ TeV. 

\subsection{Hadron colliders}
 
\begin{figure}[htb]
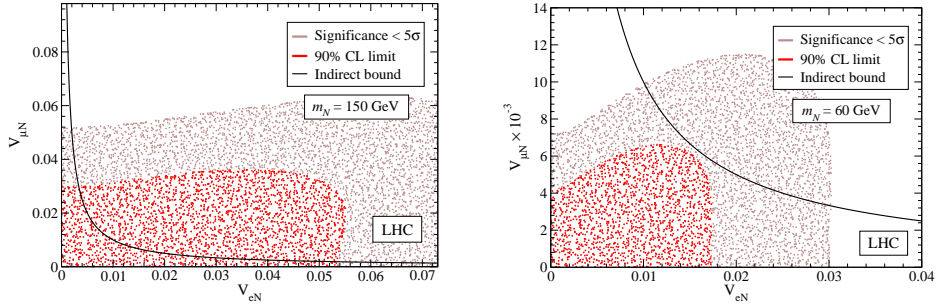

\begin{center}
\begin{tabular}{ccc}
\epsfig{file=Figs/bound-em150-LHC.eps,height=4cm,clip=} & &
\epsfig{file=Figs/bound-em60-LHC.eps,height=4cm,clip=} 
\end{tabular}
\caption{Combined limits on heavy neutrino mixings at LHC for 
$V_{\tau N} = 0$ and two heavy Majorana neutrino masses.
The meaning of the coloured areas is the same as in Fig.~\ref{fig:ilc}} 
\label{fig:lhc1}
\end{center}
\end{figure}

Hadron colliders produce large electroweak signals, and in particular
they can produce new leptons with relatively large cross sections.
If the usually huge SM backgrounds contribute relatively little to
a specific final state, one can derive non-trivial limits on these
new leptons. This is the case of heavy neutrino singlets (seesaw type I)  
\cite{Datta:1993nm,Almeida:2000pz,Han:2006ip} in like-sign
dilepton final states $\ell^\pm \ell'^\pm X$. Let us summarise the analysis of
Ref.~\cite{del Aguila:2007em}. At hadron colliders the heavy neutrino
character plays an important role because Dirac neutrinos conserve lepton number
and,
in general, their signals are overwhelmed by the backgrounds. On the other
hand, heavy Majorana neutrinos produce LNV signals,
$p \ppbar \rightarrow W^\pm \rightarrow \ell^\pm N \rightarrow 
\ell^\pm \ell'^\pm jj$, which have smaller backgrounds, and present limits
on their masses and mixings can be eventually improved.
(However, realising these masses and mixings in a specific model still
requires complicated cancellations to avoid generating too large SM
neutrino masses, as emphasised in the former section.) 
At Tevatron the signal cross sections are in practice too small, but 
at LHC they are sizeable for heavy neutrino masses of the order of the
electroweak scale (and especially for $m_N < M_W$, when the heavy neutrino
is produced on its mass shell). The limits on the mixing of a heavy Majorana 
neutrino are plotted in Fig.~\ref{fig:lhc1} for the case $V_{\tau N} = 0$
and two heavy neutrino masses above and below $M_W$, for a luminosity $L = 30$
fb$^{-1}$. We point out that for $m_N = 60$ GeV the direct limit is more
stringent than the indirect one from $\mu-e$ LFV processes.  

\begin{figure}[htb]
\begin{center}
\epsfig{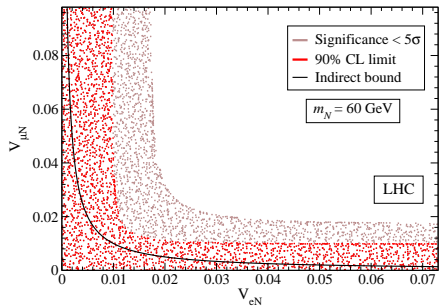}
\caption{The same as in Fig.~\ref{fig:lhc1} but for a Dirac neutrino with a
mass $m_N = 60$ GeV.}
\label{fig:lhc2}
\end{center}
\end{figure}

A Dirac neutrino does not give observable signals at LHC except if $N$
is lighter than the $W$ boson and couples to both electron and muon.
In this situation it can produce the LFV signal
$e^\pm \mu^\mp X$ with a large cross section, so that it can be observed
above the large opposite sign dilepton background. In Fig.~\ref{fig:lhc2}
we show the corresponding limits on the heavy Dirac neutrino mixings
for $m_N = 60$ GeV and $L = 30$ fb$^{-1}$.

Heavy neutrino limits improve significantly in the presence of new 
interactions, for example of a new $W'$ \cite{Keung:1983uu}
or a $Z'$ \cite{delAguila:2007ua}. In the former case, LHC is sensitive
to masses up to $M_{W'} = 3$ TeV, $m_N = 2.1$ TeV~\cite{Gninenko:2006br},
while in the latter it is sensitive to
$M_{Z'} = 2.5$ TeV, $m_N = 800$ GeV~\cite{delAguila:2007ua}
(in both cases assuming $L = 30$ fb$^{-1}$).

Finally, we point out that like-sign dilepton signals also arise in the other
two seesaw scenarios: in the production of doubly charged scalar
triplets~\cite{Hektor:2007uu}, and in pair production of fermion 
triplets~\cite{Bajc:2006ia}. For this reason,
like-sign dileptons
constitute an interesting final state in which to test seesaw at LHC.

\section{Neutrino oscillations beyond the $\nu$SM}

\begin{figure}[htb]
\begin{center}
\begin{tabular}{ccc}
\epsfig{file=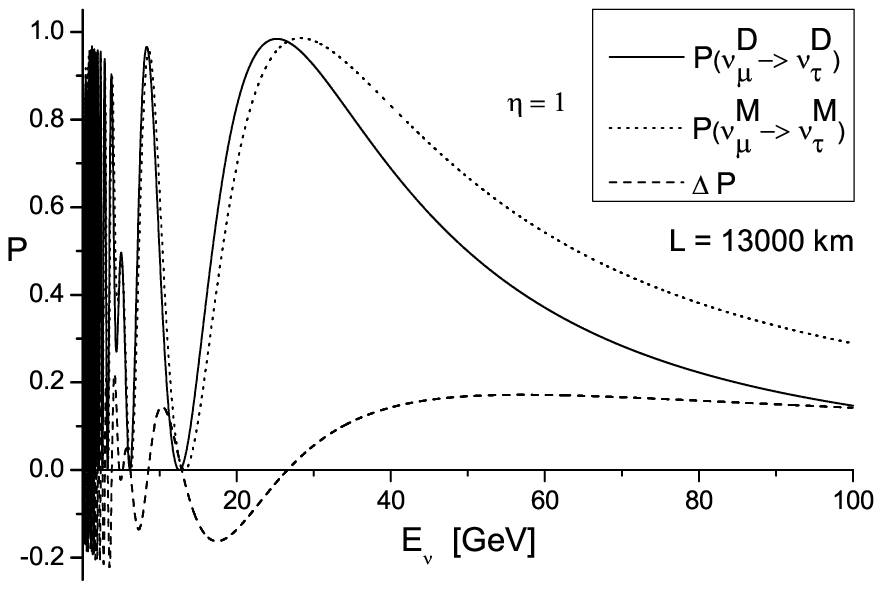,height=4cm,clip=} & &
\epsfig{file=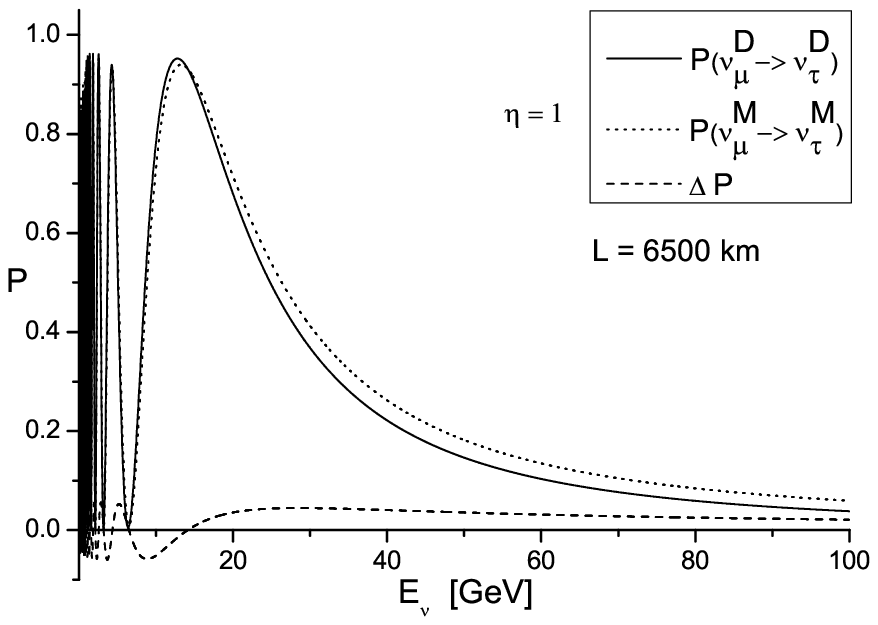,height=4cm,clip=} 
\end{tabular}
\caption{Transition probabilities for Majorana (M) and
Dirac (D) neutrinos and their difference $\Delta P$ as a 
function of the neutrino energy $E_{\nu}$ (in GeV) for 
two different baseline distances $L$. The new four-fermion interactions have a strength which is 1\% $(\eta = 1)$ of the weak interactions
\cite{Gonzalez-Garcia:2004wg}.
Note that in the Dirac case the transition amplitude with new interactions 
is the same as in the $\nu$SM.}
\label{fig:oscillations}
\end{center}
\end{figure}

Neutrino oscillation experiments will improve their precision in the future,
and they may be sensitive to new physics through its effects on light
neutrinos. For example, deviations from unitarity of the MNS matrix due to
mixing with heavy neutrinos can manifest at the percent level in
$\nu_\mu-\nu_\tau$ transitions
\cite{Bekman:2002zk}. In the presence of new right-handed interactions,
the transition probability amplitude differs if light neutrinos have Dirac
or Majorana nature, as it is shown in Fig. \ref{fig:oscillations}
\cite{delAguila:2007ug}.
The difference (dashed line) can be at the 10 \% level but only for very long
baseline distance $L$. For the examples shown, it reduces by a factor of 4
from $L = 13000$ km to $L = 6500$ km. It is also proportional to the
strength of the new four-fermion interaction.

\vspace{0.5cm}

We thank discussions with 
C. Biggio, T. Hambye, M. P\'erez-Victoria, R. Pittau and J. Syska. 
This work has been supported by MEC project FPA2006-05294,
Junta de Andaluc{\'\i}a projects FQM 101 and FQM 437, 
and by the European Community's Marie-Curie Research Training
Network under contract MRTN-CT-2006-035505 ``Tools and Precision
Calculations for Physics Discoveries at Colliders''.
J.A.A.-S. acknowledges support by a MEC Ram\'on y Cajal contract.
J.B. also thanks MEC for an FPU grant.


\begin{thebibliography}{99}

\bibitem{Yao:2006px}
  W.~M.~Yao {\it et al.}  [Particle Data Group],
  J.\ Phys.\ G {\bf 33} (2006) 1.

\bibitem{MNS}
Z. Maki, M. Nakagawa and S. Sakata, 
Prog.\ Theor.\ Phys. {\bf 28} (1962) 870.

\bibitem{Mohapatra:1998rq}
  R.~N.~Mohapatra and P.~B.~Pal,
  World Sci.\ Lect.\ Notes Phys.\  {\bf 60}, 1 (1998)
  [World Sci.\ Lect.\ Notes Phys.\  {\bf 72}, 1 (2004)].

\bibitem{Buchmuller:1985jz}
  W.~Buchmuller and D.~Wyler,
  Nucl.\ Phys.\  B {\bf 268} (1986) 621.

\bibitem{Abada:2007ux}
  A.~Abada, C.~Biggio, F.~Bonnet, M.~B.~Gavela and T.~Hambye,
  arXiv:0707.4058 [hep-ph].

\bibitem{Chen:2007dc}
  C.~S.~Chen, C.~Q.~Geng, J.~N.~Ng and J.~M.~S.~Wu,
  arXiv:0706.1964 [hep-ph].

\bibitem{Kersten:2007vk}
  J.~Kersten and A.~Y.~Smirnov,
  arXiv:0705.3221 [hep-ph].

\bibitem{delAguila:2005yi}
  F.~del Aguila, M.~Masip and J.~L.~Padilla,
  Phys.\ Lett.\  B {\bf 627} (2005) 131
  [arXiv:hep-ph/0506063].

\bibitem{Arzt:1994gp}
  C.~Arzt, M.~B.~Einhorn and J.~Wudka,
  Nucl.\ Phys.\  B {\bf 433} (1995) 41
  [arXiv:hep-ph/9405214].

\bibitem{delAguila:2005mf}
  F.~del Aguila, J.~A.~Aguilar-Saavedra, A.~Mart{\'\i}nez de la Ossa and D.~Meloni,
  Phys.\ Lett.\  B {\bf 613} (2005) 170
  [arXiv:hep-ph/0502189].

\bibitem{delAguila:2006dx}
  F.~del Aguila, J.~A.~Aguilar-Saavedra and R.~Pittau,
  J.\ Phys.\ Conf.\ Ser.\  {\bf 53} (2006) 506
  [arXiv:hep-ph/0606198].

\bibitem{delAguila:2005pf}
  F.~del Aguila and J.~A.~Aguilar-Saavedra,
  JHEP {\bf 0505} (2005) 026
  [arXiv:hep-ph/0503026].

\bibitem{Gluza:1996bz}
  J.~Gluza and M.~Zralek,
  Phys.\ Rev.\  {\bf D 55} (1997) 7030
  [arXiv:hep-ph/9612227].

\bibitem{Bergmann:1998rg}
  S.~Bergmann and A.~Kagan,
  Nucl.\ Phys.\  B {\bf 538} (1999) 368
  [arXiv:hep-ph/9803305].

\bibitem{Bekman:2002zk}
  B.~Bekman, J.~Gluza, J.~Holeczek, J.~Syska and M.~Zralek,
  Phys.\ Rev.\  D {\bf 66} (2002) 093004
  [arXiv:hep-ph/0207015].

\bibitem{Datta:1993nm}
  A.~Datta, M.~Guchait and A.~Pilaftsis,
  Phys.\ Rev.\ {\bf D 50} (1994) 3195
  [arXiv:hep-ph/9311257].

\bibitem{Almeida:2000pz}
  F.~M.~L.~Almeida, Y.~A.~Coutinho, J.~A.~Martins Simoes and M.~A.~B.~do Vale,
  Phys.\ Rev.\  {\bf D 62} (2000) 075004
  [arXiv:hep-ph/0002024].

\bibitem{Han:2006ip}
  T.~Han and B.~Zhang,
  Phys.\ Rev.\ Lett.\  {\bf 97} (2006) 171804
  [arXiv:hep-ph/0604064].

\bibitem{delAguila:2007em}
  F.~del Aguila, J.~A.~Aguilar-Saavedra and R.~Pittau,
  JHEP {\bf 10} (2007) 047
  [arXiv:hep-ph/0703261].

\bibitem{Keung:1983uu}
  W.~Y.~Keung and G.~Senjanovic,
  Phys.\ Rev.\ Lett.\  {\bf 50} (1983) 1427.



\bibitem{delAguila:2007ua}
  F.~del Aguila and J.~A.~Aguilar-Saavedra,
  arXiv:0705.4117 [hep-ph].

\bibitem{Gninenko:2006br}
  S.~N.~Gninenko, M.~M.~Kirsanov, N.~V.~Krasnikov and V.~A.~Matveev,
  CMS-NOTE-2006-098.


\bibitem{Hektor:2007uu}
  A.~Hektor, M.~Kadastik, M.~Muntel, M.~Raidal and L.~Rebane,
  arXiv:0705.1495 [hep-ph].

\bibitem{Bajc:2006ia}
  B.~Bajc and G.~Senjanovic,
  arXiv:hep-ph/0612029.

\bibitem{delAguila:2007ug}
  F.~del Aguila, J.~Syska and M.~Zralek,
  Phys.\ Rev.\  D {\bf 76} (2007) 013007
  [arXiv:hep-ph/0702182].


\bibitem{Gonzalez-Garcia:2004wg}
  M.~C.~Gonzalez-Garcia and M.~Maltoni,
  Phys.\ Rev.\  D {\bf 70}, 033010 (2004)
  [arXiv:hep-ph/0404085].


\end{thebibliography}
\end{document}